\documentclass[11pt,a4paper,english]{article}
\usepackage[T1]{fontenc}
\usepackage{amsfonts}
\usepackage[utf8]{inputenc}
\usepackage{graphicx}
\usepackage{babel}
\usepackage{blindtext}
\usepackage{color,soul}
\usepackage{amsmath}
\usepackage{csquotes}
\usepackage{subfig}
\usepackage[export]{adjustbox}
\usepackage[a4paper, portrait, margin=1in]{geometry}
\usepackage{hyperref}

\usepackage[citestyle=numeric, sorting=none]{biblatex}
\hypersetup{
    colorlinks=true,
    linkcolor=blue,
    filecolor=blue,      
    urlcolor=blue,
    citecolor=blue,
}

\title{Fear and Volatility in Digital Assets}
\author{%
    Faizaan Pervaiz \\   % \small Churchill College\\
    \small \href{https://www.cam.ac.uk}{University of Cambridge}\\
    \and
    Christopher Goh \\   % \small Hughes Hall\\
    \small \href{https://www.cam.ac.uk}{University of Cambridge}\\
    \and
    Ashley Pennington \\ % \small Trinity College\\
    \small \href{https://www.cam.ac.uk}{University of Cambridge}\\
    \and
    Samuel Holt \\
    \small \href{mailto:samuel.holt.direct@gmail.com}{samuel.holt.direct@gmail.com}
    \and
    James West\\
    \small \href{https://globedx.com}{Globe Research}\\
    \and
    Shaun Ng\\
    \small \href{https://globedx.com}{Globe Research}
}
 
\begin{document}
\maketitle

\begin{abstract} \noindent
We show Bitcoin implied volatility on a 5 minute time horizon is modestly predictable from price, volatility momentum and alternative data including sentiment and engagement. Lagged Bitcoin index price and volatility movements contribute to the model alongside Google Trends with markets responding often several hours later. The code and datasets used in this paper can be found at \url{https://github.com/Globe-Research/bitfear}.

\end{abstract}

\section{Introduction}

The volatility Bitcoin price movements consistently exceeds that of traditional securities prices including stocks, bonds, and listed futures and options on stocks and bonds.  As of March 31, 2020, Bitcoin's realised 30-day volatility reached a staggering 167 percent \cite{forbes04082020}. Realised volatilities are straightforward to calculate based on the historical price behavior of the underlying instrument itself, however, what is missing is a well-established method for calculating a forward-looking measure of volatility for Bitcoin.

There are good reasons to suspect thin-tailed Bitcoin return distributions compared to classical commodities, making the rich existing theory of options pricing viable and consequently making Bitcoin volatility easier to predict \cite{salisu2018analysing}. Regardless of Bitcoin's tail returns, it is an active area of research to price options on assets with such fat-tailed distribution \cite{taleb2019option} which will enable the pricing of Bitcoin options in either case. Alexander and Imeraj \cite{sussex91094} recently introduced VIX for Bitcoin, which along with the increasing understanding of Bitcoin's tail return characteristics and thus option pricing, makes a case for the existence of Bitcoin VIX futures, given the recent launch of Bitcoin options on the CME. 

The benchmark example of a well established volatility gauge is the S\&P 500 options market volatility index, VIX, from the Chicago Board Options Exchange \cite{exchange2003vix}. The VIX serves as the starting point for our research. There are two main uses for volatility indices like VIX: (1) to support a well functioning options market, and (2) as a gauge of the overall level of fear in the market. The S\&P 500 Index market tends to see more rapid drops and more gradual recoveries, leading to a VIX behavior that increases mainly with large market sell-offs, and decreases in stable or rising markets.  For this reason, VIX is often referred to as the fear index. %When VIX is up, the S\&P500 generally is going down.

The methodology behind the VIX, as referenced in {CBOE White Paper (2003)} \cite{exchange2003vix}, relies on gathering bid and ask quotes for options expiring between 23 and 37 days from the current point in time and reversing out the volatility implied by the options prices with a structured weighting and blending methodology to settle on an 30-day forward looking implied volatility. This approach relies entirely upon actual current or historical options prices in much the same way as backward-looking realised volatility calculations, and can be reliable when applied to a market that is deeply liquid, like the market for the S\&P500 index.

Altnerative data like Bitcoin engagement over twitter and sentiment have been previously studied by Abraham et al. \cite{abraham2018cryptocurrency}. Findings made here of the impact of non-market signals such as trend level and tweet volume on implied Bitcoin volatility may be used to enhance option (\& other derivative) pricing strategies for cryptocurrencies.

Factors such as investor sentiment are considered to be highly relevant in determining the price and volatility of Bitcoin \cite{Mendelu2016}. An aim of this paper is, therefore, to test the extent to which market sentiment has an impact on Bitcoin volatility. We continue along these lines to testing the impact of fear an emotion in Bitcoin volatility. 

Intuitively, we expect large shifts in sentiment to correlate with volatility spikes in Bitcoin prices, potentially leading the market over several hours as trading activities respond to the new information, with cryptocurrency event trading presently one of the least automated global markets. 

\section{Methodology}

For the calculation of Bitcoin volatility indices, Bitcoin index price data and bid and ask prices for Bitcoin options are required. Tweet sentiment, tweet volume, and Google Trends data is also obtained in order to test the presented hypotheses, which must then be prepared for analysis.

\subsection{Data}
\begin{figure}[hbt!]
    \centering
    \includegraphics[width=\textwidth]{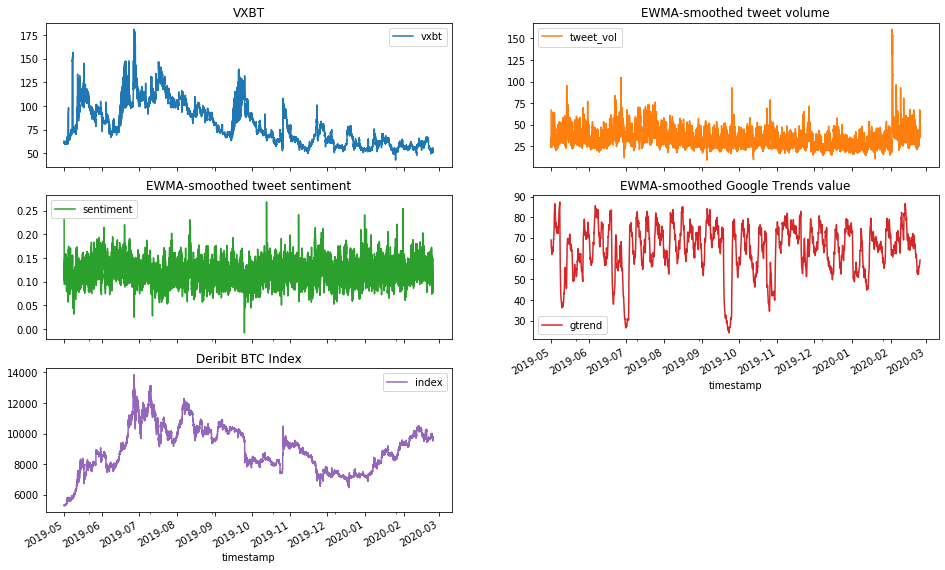}
    \caption{
        Time series data from 00:00 01 May 2019 to 00:00 25 February 2020 at 5 minute resolution
        % Bottom: Tweet volume between 06:00 25th October 2019 and 18:00 26th October 2019, with resulting tweet volume direction 
    }
    \label{fig:timeseriescombin}
\end{figure}

\subsubsection{Bitcoin options}

Deribit is a Panama-based derivatives exchange launched in June 2016. Deribit is currently one of the larger cryptocurrency options exchange by volume, offering options contracts for Bitcoin and Ethereum. The exchange lists options expiring every day, every Friday and every last Friday of the month. Trade volume is highest on the contracts expiring each Friday at 08:00 UTC, which are the contracts that will be considered in calculating a volatility index. Deribit's API does not provide a way of obtaining historical quotes, so historical data is obtained from \href{https://www.coinapi.io}{CoinAPI}. This includes bid and ask quotes for Bitcoin options from Deribit going back to 2018. With quote data collected from CoinAPI, we build a 300-day dataset of Bitcoin option quotes starting from 01 May 2019, ready for use in volatility index calculations.

\subsubsection{Twitter}

A Python scraper utilising the \textit{tweepy} library continuously downloads and stores tweets containing the hashtag \#Bitcoin in real time. Only original tweets are recorded; retweets as defined by the Twitter API and tweets containing the term `RT' are discarded to prevent sentiment values being skewed by repeated content instead of measuring the sentiment of original content.

To obtain historical tweets, the scraping library \textit{GetOldTweets3} is used to again retrieve all tweets within the given time period containing the hashtag \#Bitcoin, again discarding retweets. Combining tweets collected live and by historical search we build a dataset of all original \#Bitcoin tweets from 01 May 2019 to 25 February 2020. This dataset spans 300 days and contains 2.02 million tweets.

\subsubsection{Google Trends}

Using the \textit{pytrends} library we retrieve Google Trends data for the term `Bitcoin'. The data consists of an hourly trend value time series, where the trend value reflects the search interest relative to the highest during the given period in time. A value of 100 indicates peak interest in the term, whilst a value of 25, for example, means interest in the term is 25\% of peak interest.

\subsection{Time Series Construction}

The time series generated in the following sections that go on to becoming inputs to our model are shown together in Figure~\ref{fig:timeseriescombin}.
\subsubsection{VIX \& VXBT}

VIX is a `fear gauge' used to measure the market’s expected volatility implied by at-the-money option prices \cite{exchange2003vix}.

To calculate an implied volatility time series from Bitcoin options data, we utilise CBOE\'s VIX method, assuming a risk-free rate ($R$) of zero \cite{sussex91094}. This index is labelled VXBT. The contribution of each expiry to the index is defined as such:
\begin{align}
\sigma^2 = \dfrac{2}{T}\sum_i \dfrac{\Delta K_i}{K^2_i}e^{RT} Q(K_i) - \dfrac{1}{T}\Big(\dfrac{F}{K_0}-1\Big)^2
\end{align}
Where $K_i$ is the strike price of the i-th out-of-the-money option, $\Delta K_i$ is half the difference between the strike prices on either side of $K_i$, F is the strike prices of the option with the minimum absolute difference between call and put prices with the sum of that call price minus that put price, $Q(K_i)$ is the midpoint of the bid-ask spread for each option with strike $K_i$, $K_0$ is the strike price equal to or otherwise immediately below F and T is the time from the option to expiry in minutes. 

Using this formula, we interpolate the VXBT with the near term options contribution $\sigma^2_1$ with time to expiry $T_1$ and the next term options contribution $\sigma^2_2$ with time to expiry $T_2$:
\begin{align}
\text{VXBT} = 100 \times \sqrt{\Big\{T_1 \sigma^2_1\Big[\dfrac{N_{T_2} - N_7}{N_{T_2} - N_{T_1}}\Big] - T_2
\sigma^2_2\Big[\dfrac{N_7 - N_{T_1}}{N_{T_2} - N_{T_1}}\Big]\Big\} \times \dfrac{N_{365}}{N_7}}
\end{align}

VXBT differs from VIX in that rather than taking options expiring between 23 and 37 days from now, the options expiring on the next two Fridays are considered, giving the index a maturity of 7 days as opposed to VIX's 30-day maturity.

With this method we calculate VXBT at five-minute intervals from midnight on 01 May 2019 to midnight on 25 February 2020.

\subsubsection{Twitter data}
Twitter data is used to develop two time-series: tweet sentiment and tweet volume.

Each tweet is assigned a compound sentiment score by the VADER (Valence Aware Dictionary and sEntiment Reasoner - \cite{hutto2014vader}) tool. VADER calculates a positivity, neutrality and negativity score for each text input and the compound score is a normalised combination of these scores. 

Tweets are grouped by their time of creation into five-minute intervals, which then allows us to create a time series of tweet volume. On initial inspection there appears to be weekly cycle seasonality in tweet volume over time. The mean compound sentiment score is calculated for each five-minute interval to create a tweet sentiment time series. An exponentially weighted moving average (EWMA) with a window size of 12 (i.e. looking back up to an hour in the past) is applied to both time series for smoothing and to place greater emphasis on recent changes in volume/sentiment over distant fluctuations. Data leakage is avoided as only past observations are used to calculate the EWMA.

\subsubsection{Google Trends data}
Google Trends value data obtained for the term 'Bitcoin' is upsampled with linear interpolation from one-hour intervals to five-minute intervals, to match the sample rate of the tweet sentiment, tweet volume and VXBT time series.

\subsection{Classification and Permutation Importance}

After applying min-max normalisation to each time-series, we follow Tsantekidis et al. \cite{tsantekidis2017forecasting} to assign a direction value of 1, 0 or -1 at each time for each time series based on whether the value of that time series has increased above a percentage threshold, stayed the same or decreased below the percentage threshold since its value five minutes earlier . The percentage threshold value for each time series is found by a greedy algorithm that aims to equalise the number of values in each class 1, 0 and -1. This forms a classification problem where the direction of values of tweet sentiment, tweet volume and Google Trends data can be used to predict the future direction of Bitcoin volatility VXBT. 

The training input is constructed by taking a window of the previous 24 observations of each feature to predict the output, which is the direction of VXBT during the next time step. In other words, for data with time interval $T$, at current time $t$ we take the directions of tweet volume, tweet sentiment, Google Trends data and VXBT at times $t, t - nT, t - 2nT, ..., t - 24nT$ to predict the direction that VXBT will have taken by time $t + nT$. In this case, $T$ equals five minutes, meaning the window length is 2 hours. A simple train-test split of 90-10 is applied to the input and output: the first 270 days of data are used to train the model, with the remaining used for testing.

After training various classification models on the dataset, we can use permutation importance to determine the features with the most predictive power for VXBT. Permutation importance is a method of calculating feature importance by shuffling the values of each input feature to the model and calculating the resulting drop in model accuracy. A large drop in accuracy indicates an important feature while a small drop indicates a less important feature. 

 \subsubsection{Gradient Boosting Classification (GBC)}
 Gradient boosting classification is an additive model that works by taking many weak predictive models, such as a single decision tree, and continually building upon the output of these models, adding a new weak predictive model at each stage that is based on the output of all the previous models \cite{gradientboosting}. The summation of all these weak predictive models that build upon the mistakes of previous models is what gives gradient boosting classification its name.
 
 In a gradient boosting classifier, we start with an initial estimate on what the predicted outcome should be. The residual, which is the difference between our initial estimate and the data, is used as the dependent variable in building a new predictive model. The sum of our initial estimate with the output of all the predictive model we have built multiplied by their individual learning rate coefficients gives us a new estimate. Further residuals are generated by taking the difference between this new estimate and the data. Our model is trained when we have reached the number of predictive models we want to add or the residual cannot be improved further.
 
\begin{figure}[!htb]
    \centering
    \subfloat[Confusion matrix]{\includegraphics[width=0.5\textwidth,valign=c]{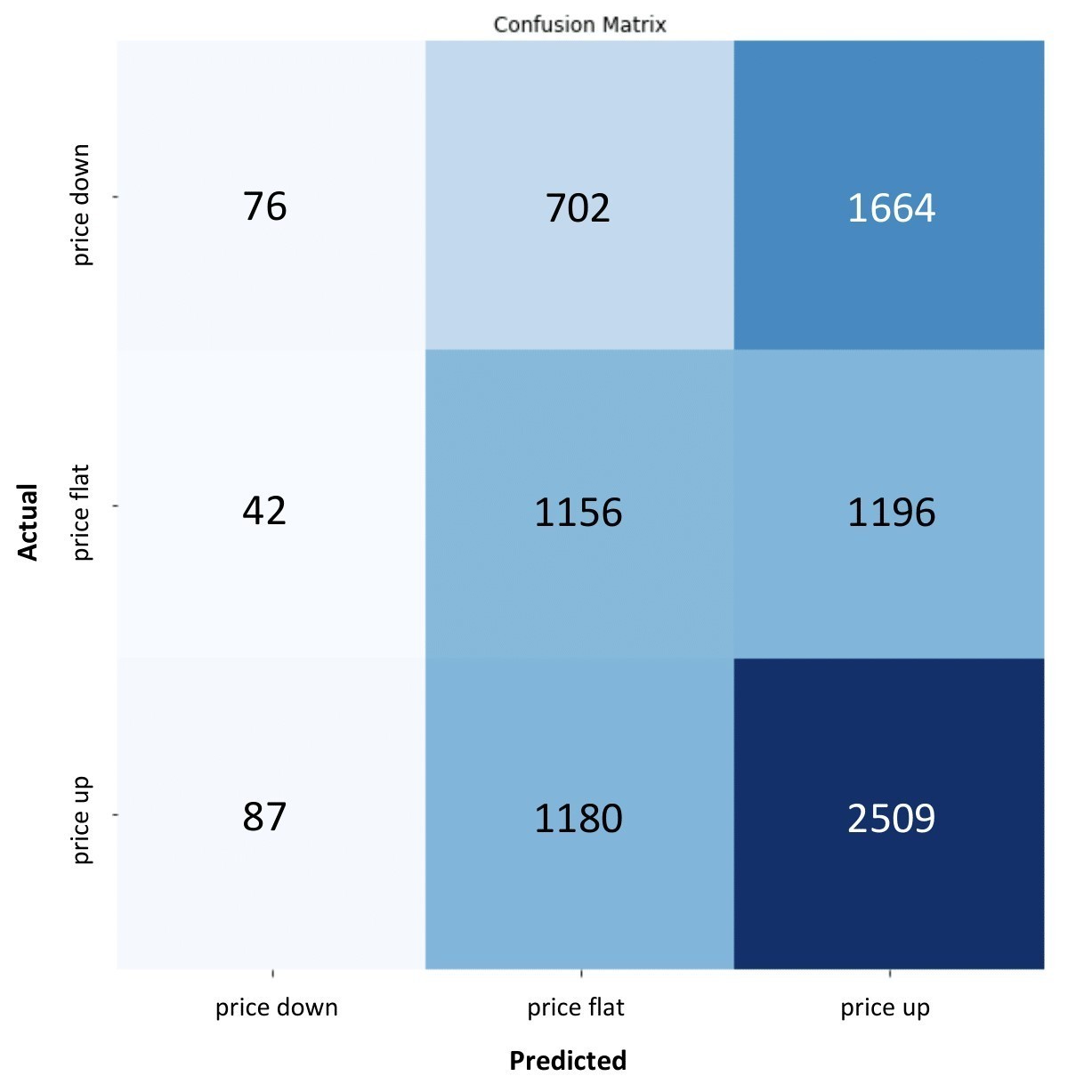}}
    \hfill
    \subfloat[Permutation importance scores for top 20 features of: all features (left) and non-financial features only (right)]{\includegraphics[width=0.5\textwidth,valign=c]{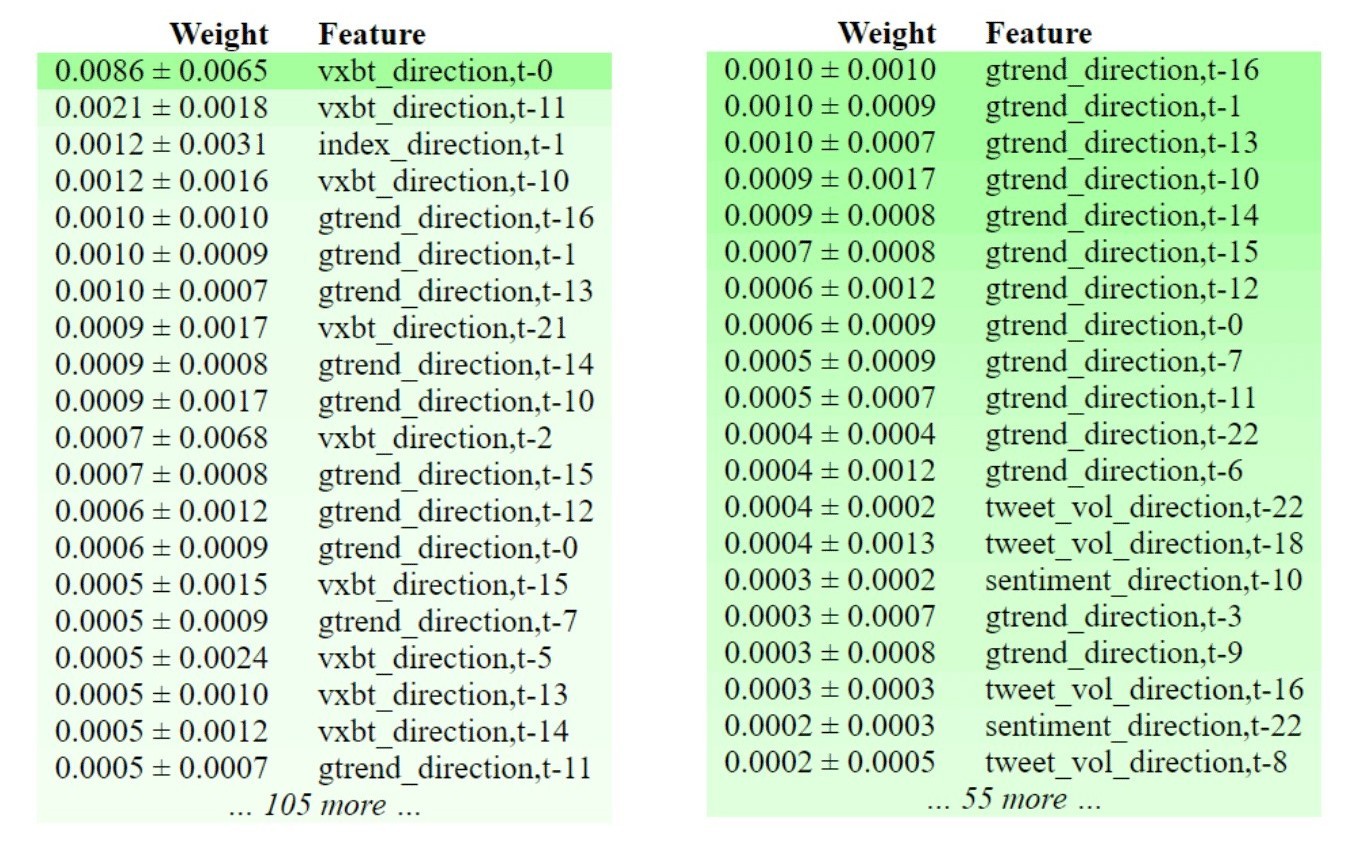}}
    \caption{Confusion matrix and permutation importance scores for tuned gradient boosting classifier}
    \label{fig:gbc_mv_cm}
\end{figure}

A baseline gradient boosting classifier is trained on the training set with five-fold cross validation. This is followed by a random search where each of the hyperparameters learning rate, number of regression stages and number of estimators is randomly taken from a predefined range and the resulting accuracy score calculated. The hyperparameter values for the best performing model are then used to reduce the predefined ranges for a grid search to find optimal hyperparameter values. The final accuracy, confusion matrix and permutation importances are given in Figure~\ref{fig:gbc_mv_cm}.

\section{Results}

\subsection{Volatility momentum and search trends are predictive of future volatility}

We follow along the lines of de Prado 2020 \cite{ClusteredFeatureImportance2020}, using permutation importance as a means of determining feature relevance in historic time series data. Applying this across our 5 minute resolution, 300 day dataset, we find overall that \label{section:modeldiscussion} the tuned GBC achieves a test set accuracy of 43.4\%, considerably better than random chance for a 3-outcome problem (random score of 33\%). As discussed previously, the test set contains time steps at five-minute intervals over a period of 30 days, where each step consists of 120 inputs - the directions of each of the five features over the last 25 time steps. The training set is structured in the same way with data from the 270 days preceding the start of test set data.

The most important feature in determining the direction of VXBT in the next five minutes is found to be the direction of the VXBT in the previous five minutes. This is somewhat to be expected as any drawn out increases or decreases in the volatility value will result in periods where the current direction is the same as the previous direction.

The short term price momentum i.e. direction of index price five minutes prior ($index\_direction,t-1$) also has high importance. It is unsurprising that index direction has high importance as the VXBT calculation is based on out-of-the-money puts and calls, the quantities and inclusion of which are directly affected by the current price index level.

It is highly encouraging to find that half of the top 20 features are from non-financial indicators. In fact, all of these are Google Trends value directions, which immediately suggests this is a strong predictor for future VXBT directions. The time distance of these indicators ranges from $t-0$ to $t-16$, i.e. current time to 80 minutes earlier. Looking at the top 20 non-financial indicators shows that Google Trends dominates, making up 14 of the 20 and all of the top 12. Additionally, three of the top 12 Google Trends features are within the previous 30 minutes, indicating to some extent that Google Trends leads VXBT volatility very closely.

In contrast to Abraham et al. \cite{abraham2018cryptocurrency} we find weak evidence for tweet volume having predictive power here. It is possible that since we have constructed our model at a much higher sample rate, significant day-to-day variations in tweet volume that have high predictive power have been lost to small and perhaps cyclical hourly variations with little predictive power.

Other than the low importances of tweet volume features, our results agree with those of Abraham et al.: Google Trends data is found to have strong predictive power, while sentiment, due to general invariance of tweet sentiment regardless of market events, has little to no predictive power. The result is Google Trends values appear to have strong predictive power for both cryptocurrency prices and the volatility index VXBT.

\subsection{Roughly 1 hour lag in pricing in new information}

Whilst the previous 5-10 minutes of volatility momentum and google trend interest accounts for a large amount of our model performance, we expect a slightly longer historic lookback period to be predictive of intraday volatility movements. 

\begin{figure}
    \centering
    \includegraphics[width=0.7\textwidth, height = 8 cm]{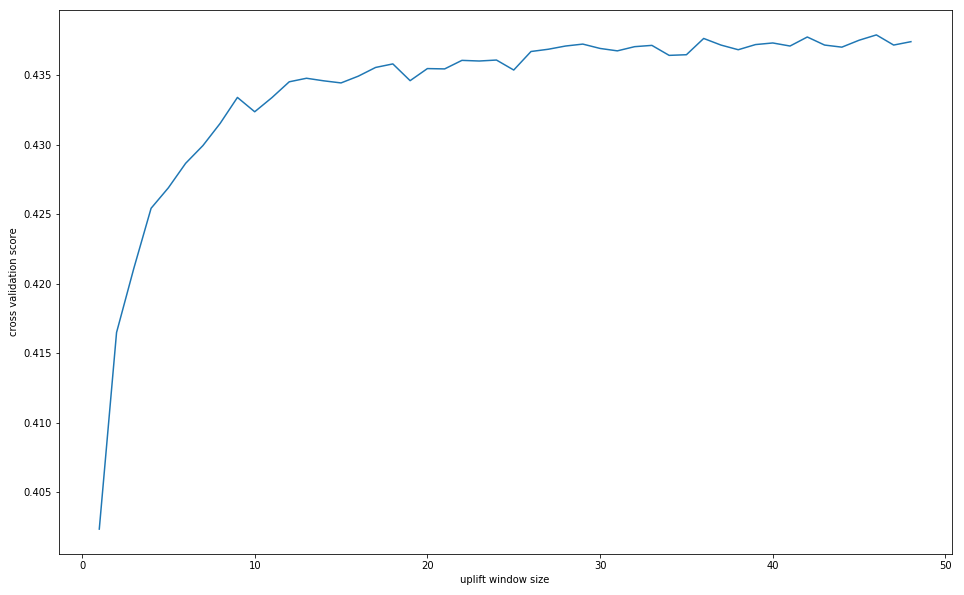}
    \caption{Model performance as a function of historic window size used (5 minute intervals)}
    \label{fig:uplift}
\end{figure}

To measure this we compare cross-validated model performance across window sizes from 5 minutes up to 4 hours, as per Figure~\ref{fig:uplift}. Performance is roughly increasing as expected with a longer window, saturating after around an hour or so. We interpret this as the approximate rate at which new information about the future volatility of the Bitcoin exchange rate is priced into the market, rather than a consequence of the modelling process. 

\subsection{Alternative data can build a richer fear gauge}
Considering all of the above, it is possible for a generalised fear index for Bitcoin to be constructed. VXBT is the de facto financial fear gauge for Bitcoin, and Google Trends values have strong predictive power not just for Bitcoin prices but also for VXBT, making it a strong `fear component' to be included in this index. Exploring tweet volume further may provide a further component to this index.

\section{Discussion}
By collecting a dataset of Bitcoin options prices over a 300 day period and building an implementation of CBOE's VIX, we have constructed a time series of the forward-looking volatility index VXBT with seven-day maturity. After combining this with tweet sentiment, tweet volume and Google Trends data and discretising based on the direction of price movements within a given time interval, we trained a gradient boosting classifier model to determine the predictive power that these features have for VXBT.

We find that Google Trends values hold by far the highest and most reliable importance out of all the non-financial indicators investigated in this paper, with some evidence to show that this indicator leads the VXBT. This contradicts our initial hypotheses and expectations, namely that Google Trends values would lag VXBT; however it would seem that the phenomenon theorised for tweet volume also holds for Google Trends values - investors and traders use search engines to research Bitcoin-related external events affecting their position and subsequently make trades that influence the cryptocurrency's volatility. However, further scrutiny is required before concluding there is a direct link as there is a possibility this result is affected by data leakage. Google Trends values are normalised and not absolute, and due to API limitations, Trends value data with hourly resolution can only be obtained for a week at a time. This means to generate a continuous scaled time series that spans 9 months like in our case, normalisation is applied to the entire set using the range of values from each week. Future values therefore impact current Trends values and studying the extent to which this leakage skews this result would be worthwhile.

Further research is needed to determine the reason for the lack of predictive power of tweet volume. As previously suggested, a five-minute resolution may simply be too high to capture the longer term trends in tweet volume that have predictive power for cryptocurrency prices and potentially VXBT.

Further studies could also include other non-financial indicators. Sentiment from articles featured on the aggregator Google News could be used as an indicator as news articles perhaps have more sentiment bias than tweets and less noise, making it potentially a more useful feature than tweet sentiment. Other aggregator platforms such as Reddit could also be explored. Subreddits such as r/Bitcoin are expected to have more focused, sentiment heavy conversations compared to Twitter, however there may be a strong positive bias due to `HODLing' - buying and holding regardless of price. We could also explore the components of a pre-existing generalised fear index such as the one available at \url{https://alternative.me/crypto/fear-and-greed-index/}. Alongside Trends and social media sentiment, indicators such as historical volatility, dominance and market momentum are included. It is worth investigating the usefulness of these indicators in our own generalised fear index.

% There is also scope for building a volatility index implementation more suited to Bitcoin. In calculating VXBT, we replicate the VIX calculation \cite{exchange2003vix} during which out of the money calls and puts are found by separating at a strike price equal to or immediately below the strike price where the difference between the option bid and ask prices is minimum (the forward price). Figure~\ref{fig:vix_jumps} shows how this results in non-ideal jumps in the value of VXBT as a result of sudden changes in this separation strike price due to option price movements on Deribit This effect was allowed in our analysis as a feature of the calculation and the Bitcoin market, however adapting the VIX calculation to cope with these strike price jumps may provide a more useful fear gauge and forward looking volatility index. This may be done using an adaption of the VIX Index Filtering Algorithm recently developed by CBOE. The suggested AVXBT \cite{sussex91094} does not correct these jumps.

%\begin{figure}[hbt!]
%    \centering
%    \includegraphics[width=\textwidth]{figs/vxbt_jumps.png}
%    \caption{VXBT jumps caused by jumps in separation strikes K0,1 (near term) and K0,2 (next term)}
%    \label{fig:vix_jumps}
%\end{figure}

\section{Acknowledgements}

The research in this paper was made possible by resources provided by Globe Research as part of Globe, a pioneering cryptocurrency derivatives exchange, available at \url{https://globedx.com}.  The code and data used in this project can be found at \url{https://github.com/Globe-Research/bitfear}.

\printbibliography

@article{sussex91094,
    month = {April},
    title = {The Bitcoin VIX its variance risk premium},
    author = {Carol Alexander and Arben Imeraj},
    publisher = {Portfolio Management Research},
    year = {2020},
    journal = {Journal of Alternative Investments},
    url = {http://sro.sussex.ac.uk/id/eprint/91094/}
}

@inproceedings{hutto2014vader,
  title={Vader: A parsimonious rule-based model for sentiment analysis of social media text},
  author={Hutto, Clayton J and Gilbert, Eric},
  booktitle={Eighth international AAAI conference on weblogs and social media},
  year={2014}
}

@article{exchange2003vix,
 title = {Vix white paper},
 author ={Exchange, Chicago Board Options},
 url = {http://www.cboe.com/micro/vix/vixwhite.pdf},
 year = {2003}
}

@article{Mendelu2016,
  title={Sentiment and Bitcoin Volatility},
  author={Jaroslav Bukovina and Matúš Martiček},
  url={https://pdfs.semanticscholar.org/e2b7/41c747e9ba55be43c35e9592e85c1caee661.pdf},
  year={2016}
}

@article{forbes04082020,
 title = {Bitcoin Volatility Reached A 6-Year High In March},
 author ={Charles Bovaird},
 url = {https://www.forbes.com/sites/cbovaird/2020/04/08/bitcoin-volatility-reached-a-6-year-high-in-march/},
 year = {2020}
}

@article{ClusteredFeatureImportance2020,
 title = {Clustered Feature Importance},
 author ={Marcos Lopez de Prado},
 url = {https://ssrn.com/abstract=3517595},
 year = {2016},
 month = {1}
}

@article{gradientboosting,
author = {Natekin, Alexey and Knoll, Alois},
year = {2013},
month = {12},
pages = {21},
title = {Gradient Boosting Machines, A Tutorial},
volume = {7},
journal = {Frontiers in neurorobotics},
doi = {10.3389/fnbot.2013.00021}
}

@inproceedings{tsantekidis2017forecasting,
  title={Forecasting stock prices from the limit order book using convolutional neural networks},
  author={Tsantekidis, Avraam and Passalis, Nikolaos and Tefas, Anastasios and Kanniainen, Juho and Gabbouj, Moncef and Iosifidis, Alexandros},
  booktitle={2017 IEEE 19th Conference on Business Informatics (CBI)},
  volume={1},
  pages={7--12},
  year={2017},
  organization={IEEE}
}

@article{abraham2018cryptocurrency,
  title={Cryptocurrency price prediction using tweet volumes and sentiment analysis},
  author={Abraham, Jethin and Higdon, Daniel and Nelson, John and Ibarra, Juan},
  journal={SMU Data Science Review},
  volume={1},
  number={3},
  pages={1},
  year={2018}
}

@article{taleb2019option,
  title={Option Pricing Under Power Laws: A Robust Heuristic},
  author={Taleb, Nassim Nicholas and Yarckin, Brandon and Mann, Chitpuneet and Delic, Damir and Spitznage, Mark},
  journal={arXiv preprint arXiv:1908.02347},
  year={2019}
}

@techreport{salisu2018analysing,
  title={Analysing the distribution properties of Bitcoin returns},
  author={Salisu, Afees A and Tiwari, Aviral Kumar and Raheem, Ibrahim D and others},
  year={2018}
}

\end{document}